\documentclass[12pt,a4paper]{article}
\usepackage{mathtext}
\usepackage[cp866]{inputenc}
\usepackage{epsfig}
\countdef\pageno=0 \pageno=1
\begin{document}

\title{ The  Knee in Galactic Cosmic Ray Spectrum and Variety in
 Supernovae}

   \author{ Lyubov G. Sveshnikova}

   \date{}


   \maketitle

\begin{center}
              {\it Skobeltsyn Institute of Nuclear Physics of Moscow State 
 University,\\ 119992, Leninskie Gory, MSU, Moscow,
 Russsia. 
 email: sws@dec1.sinp.msu.ru}

\end{center}

{\small
\abstract{
 Here is proposed the qualitative and quantitative solution
of long-standing problem in astrophysics: the origin of 'knee' in 
Galactic cosmic rays (CRs). The paper deals with calculations based on the 
standard model of cosmic ray acceleration in Supernova remnants (SNR)
and on the latest astronomical data concerning the variety in Supernovae
types and energies as well as the sites of their explosions.
The results obtained allow explanation of the main features of all particle
cosmic ray spectrum: the  'knee' around 3 PeV, the change of a slope
 by 0.3-0.4 in this point and the start of dip around 1000 PeV.
The results are
based essentially on the premise that energy of accelerated CRs
in the given SNR is proportional to the kinetic energy of this explosion,
that  stresses the input of high energy explosions and suppresses 
the contribution of low energy explosions to the total CR flux. } 
 }

\small
\section{Introduction}    
	    Supernovae (SNe) represent the catastrophic explosions that mark
the end of the  life of some stars. It is well known, that the mechanical
energy input  to the Galaxy from each supernova is about $10^{51}$ erg,
so with a rate 
of about 0.01-0.03 year$^{-1}$ total power of SNs in our Galaxy is enough
to provide  the total energy of Galactic cosmic rays (GCR)
 $\sim 10^{-12}$ erg/cm$^3$
(Berezinsky et al. \cite{berez}). It is shown that there exists a
 mechanism needed for the channeling about
$10\%$ (or even more (Berezhko et al. \cite{berezh2000a})
 of the mechanical energy of
the explosion into relativistic particles.
 Considerable collective efforts have
been made during the recent years to clarify the  mechanism of  CR
acceleration in SRs (Drury et al. \cite{drury}). Theoretical
progress is connected  with the development of the
kinetic nonlinear theory of diffusive shock  acceleration
(Berezhko et al. \cite{berezh1996}, \cite{berezh1997}, 
 \cite{berezh2000a}, \cite{berezh2000b}; Berezhko$\&$ Ellison \cite{berezh1999}; 
Ellison et al. \cite{ellis1997}, \cite{ellis2000}; Drury et al. \cite{drury};
 Malkov et al. \cite{malko}) and key
advance has been made due to improved understanding of the nonlinear reaction
effects on the shock structure. This theory is able
 to explain not only
the main  characteristics of the observed all particle cosmic ray spectrum
 up to an energy of
$10^{14}-4\cdot 10^{14}$ eV 
(Berezhko et al. \cite{berezh2000a};  Drury et al. \cite{drury}),
  but also  heavy element abundances in CR flux
relative to the solar system (Ellison et al. \cite{ellis1997}).  

        The standard model predicts (Drury et al. \cite{drury}):

    1) The power-like and approximately similar spectra of various nuclei
       of CRs beyond an energy of 100 GeV/n. Spectra of heavy component can be
       slightly harder ($\Delta  \gamma \sim 0.3$) than proton one
       due to more effective acceleration of dust grains and ions
       (Ellison et al. \cite{ellis1997}). 

    2) The slope of power-like  source spectrum is $\gamma_{sour}\sim 2.0-2.1.$
       May be  spectra have a 'curvature' being even  harder before the
       maximal energy of acceleration, if nonlinear reaction effects are strong
       (Berezhko et al. \cite{berezh2000a}), (Ellison et al. \cite{ellis1997}).
        
    3) The maximal energy of accelerated particles $E_{max}$ (cut off energy) is 
       $\sim Z \times 10^{14}$ ev for the average  SNe exploding into the average
       interstellar medium (ISM).
 
    4) There is a possibility to move the maximal energy to higher energies
       assuming nonordinary medium, for example, circumstellar medium  
       (CSM) for any class of explosions: explosions into wind of Wolf-Rayet
       stars or  explosions into superbubbles (Bykov $\&$ Toptygin \cite{bykov1997}),
       see also reviews
       (Ptuskin \cite{ptusk}; Biermann \cite{bierm}).
       This effect is connected   mainly with the higher magnetic field in the
       stellar or superbubble interior.  
     
    5) The actual source spectra inferred from observations after propagation
       corrections is $\gamma_{sour}=\gamma_{obs}-\Delta \gamma $. 
       The value of $\Delta \gamma$=0.6 for simple
       diffusion model and $\Delta \gamma$=0.3 for model with reacceleration, 
       $\Delta \gamma$=0.8 for model with Galactic wind (Jones et al. \cite{jones}).
       So the value of $\Delta \gamma$ is not well known yet.
    
       The most 'nasty' problem, as it was called in (Drury et al. \cite{drury}),
 is the knee problem.
The standard picture makes a clear prediction, that the GCR spectrum
should start to cut off at energy about $10^{14}$ eV or less for all species
and drop exponentially as one goes to higher energies. Only some subclass
of SNR can provide the knee particles while most SNRs have spectra cutting
off at considerably lower energies (Reynolds $\&$ Keohane \cite{reyno}).

       The upper limit of accelerations is determined essentially
 by the product
 of shock radius $Rsk$, shock velocity $Vsk$, ejected mass $Mej$, remnant
 age $Tsnr$,
 explosion energy $E_{51}$, usually normalized to value of $10^{51}$ erg.
 All these
 values are connected with each  other and vary from  explosion to  explosion. 
 $E_{max}$
 can be expressed by the simple formula, if only Sedov phase of SNR
 expansion is considered (Ellison et al. \cite{ellis1997}):
        
$$ (\frac{E}{A})_{max}=2\cdot 10^{14}\left(\frac{Z}{A}\right)
\left(\frac{0.3\cdot B}{\eta \cdot 3\mathrm{\mu G}}\right)
\left(\frac{n_H}{\mathrm{cm^3}} \right)^{-1/3}
\left(\frac{E_{SN}}{10^{51}\mathrm{erg}}\right)^{1/3}
\left(\frac{Vsk}{10^3 \mathrm{km/s}}\right)^{1/3}~\mathrm{eV} \eqno (1).$$

       A week dependence on $Vsk$ and 
on $E_{51}=E_{SN}/10^{51}$ erg,  but
strong dependence on magnetic field $B$ of ISM can be seen there.
 This formula predicts
$E_{max} \sim 100\cdot Z$ TeV
for $E_{51}=1$  and for the density of protons in medium
 $n_H\sim 0.3$ cm$^{-3}$.
An accurate determination of $E_{max}$ in frames of a more complete kinetic theory
(Berezhko et al. \cite{berezh2000a}, \cite{berezh2000b}) taking into account timescale evolution of the shock
 in both Sedov phase and  in free expansion phase of SNR evolution, gives,
within a factor of 2, the same results (Drury, \cite{drury}). 
 
The main idea of this work (it seems very natural) is an attempt
 to get the average cosmic ray particle spectrum
taking into account the distribution of SNRs in types, explosion energies,
type of interstellar and circumstellar medium around it. In this case
the total CR flux F(E) can be expressed by the formula 

$$    F(E)=\sum_{i=1}^{Nz} \sum_{j=1}^{Ntp}
 \int\limits_{E_{51}min}^{E_{51}max} \Psi_j (E_{51})
 G(E, E_{max}(E_{51}),B,Z) dE_{51}, \eqno(2)$$ 

where $\sum_i$ is the summation by different  cosmic ray nucleus groups Nz, 
$\sum_j$  is the summation by different types of SNe explosions Ntp.
$G(E,E_{max}(E_{51}, B, Z))$ is the average spectrum
 of comic rays in every explosion,
which
was approximated by the power low:

$$G(E,E_{max})=Io E^{-\gamma}, \eqno(3) $$ 

$$\gamma=2.0~ \mathrm{in~ the~ interval} ~ 10~GeV < E < E_{max}/5;$$\\
$$\gamma=1.70~  \mathrm{in~ the~ interval}~ E_{max}/5 < E < E_{max};$$\\
$$\gamma=5~\mathrm{in~ the~ interval}~ E > E_{max}.$$

This spectrum shape  takes into account nonlinear  reaction of CRs on
shock structure (Berezhko et al. \cite{berezh1997}, \cite{berezh2000a}):
the decreasing of $\gamma$ before $E_{max}$.
$E_{max}$ depends on $E_{51},~ Vsk,~ B,~ n_H$ accordingly to formula (1). 
For every type of explosion the value of $B$ and $n_H$ can be different.

Intensity of CRs produced in every SNR is found from the condition
$$\int G(E, E_{max})E dE = 0.1\cdot E_{51}. \eqno(4)$$ 

As it was shown in (Berezhko  \cite{berezh2000a}) the portion
of SNR kinetic energy transforrmed to CRs can reach even $30\%$.

The most indefinite distribution is the distribution in $E_{51}$
 for various types
of SNe. The third  section  will be devoted to this problem. In the second
section   the latest astronomical observational data about variety in 
SNe explosions will be reviewed. In the fourth section we present
the numerical results of
 calculations. In the fifth part of the paper we discuss physical interpretation
of the results obtained.  

But we would like to note in advance that the diversity of SNe by
explosion energies  results in very 
unexpected and promising conclusion: the knee region moves to higher
energy by a factor about of 5-10, though we use only the standard model of CR
acceleration and add to it
the latest astronomical data on Supernovae explosions.

\section{Variety in Supernovae}

  Detailed observations of a growing number of Supernovae show that
the nature of this phenomena is complex (Turatto \cite{turat2003}), many new peculiar events
discovered during last years display the wide range of luminosities,
expansion velocities, chemical abundances, that is evidence for large
variations in explosion energy and in the properties of their
progenitors (Hamuy \cite{hamuy}). 

The thermonuclear explosions of accreting
white dwarfs that explode as they approach the Chandrasecar mass 
($\sim 1.4 M\odot$)  produce type Ia SNe, which owing to their high luminosity
and accurate calibration are successfully used for the determining
the geometry of Universe (Leibundgut \cite{leibu}), as a 'standard candles'.
The energies of  the explosions are practically fixed.

      Core collapse supernovae CCSNe (SNIb/c, SNII) are thought
to be the gravitational collapse of massive stars ($M > 8 M\odot$), which
makes  the  neutron stars as a compact remnants.  CCSNe prove to comprise
the most common general class of exploding stars in the Universe and they
come in a great variety of flavors (Hamuy \cite{hamuy}). Even subclass of 'normal'
SNII  : plateau SNII-P and Linear SNII-L demonstrates a wide range
in explosion energy, from 0.6 to 5.5 ($\times 10^{51}$) erg among classical SNII (Hamuy, 2003).
The ejected masses also are in broad range between 14 and 56 $M \odot$ 
 (Hamuy \cite{hamuy}). Despite the great diversity displaced by SNII-P, these objects
show a tight luminosity-velocity correlation. This result suggests that, while
 the explosion energy increases so do the kinetic energies  (Hamuy \cite{hamuy}).
These stars explode as  isolated stars.

       A distinct class of SNIIdw can be identified which are believed
to be strongly interacting with a 'dense wind' produced by SN progenitors
 prior explosion. These SNe have strong radio emission caused by the
 interaction with circumstellar medium (Chevalier \cite{cheva}) (mass loss rate is
 $\sim 10^{-4} M \odot$) for their progenitors. SNIIdw comprise $\sim 30\%$
 of all CCSNe (Hamuy \cite{hamuy}). When the narrow line is present, the SN is
 classified as IIn  ('narrow'). The strong degree of individuality
 is seen in their spectra, but
 despite the great photometric diversity among SNIIdw, these objects share the
 property of being generally more luminous than the classical SNII 
 (Hamuy \cite{hamuy}). Among  this type SNe one event SN 1997cy is much more
 energetic than any other
 SNII ($E\sim 30 \cdot 10^{51}$ erg (Turatto \cite{turat2000})).
 SN 1997 cy and its twin SN 1999E (Rigon \cite{rigon})
 are associated to GRBs. As in the case
 of others these events show strong ejecta-CSM interaction with explosion
 energies as high as $3\cdot 10^{52}$ ergs (Turatto \cite{turat2003}).

       Hydrogen-deficient supernovae  SNIb and SNIc are associated
with the gravitational collapse of massive stars (may be Wolf-Raet stars),
which have lost their
hydrogen envelope during the phase of strong wind. In the case of SNIc
most of the helium is gone as well. There is yet no direct observational
proof for binary companions in SNIb/c, but this seems likely (Turatto \cite{turat2002}). 

In the past
few year 3 SNe (1997ef, SN 1998bw, SN 2002 ap) have been
found to display very particular spectra: they are extremely  smooth
and featureless, that can be interpreted as the result of unusual expansion
velocities (Hamuy \cite{hamuy}). This fact suggests that these objects are
hyper-energetic so they are called  'hypernovae'. 
The estimated energies of explosions are very high: 
$7\cdot 10^{51}$ for 2002 ap (Mazzaly \cite{mazza}),
$8\cdot 10^{51}$ for 1997ef (Nomoto et al. \cite{nomot2000}),
$60\cdot 10^{51}$ for 1998 bw (Nomoto et al. \cite{nomot2000}).
The estimated expansion velocity of this object is as high as 
$> 30000$ km/s (Turatto et al. \cite{turat2002}).
SNe 1998bw was not only
remarkable for its great expansion velocity and luminosity, but
also because it exploded at nearly the same location and time as 
GRB980425 (Galama et al. 1998).

 M. Hamuy in the review (Hamuy \cite{hamuy}) makes the following conclusions.
Despite the great diversity of core collapse SNe several regularities
emerge which suggest that 1) there is a continuum in the properties of these
objects, 2) the mass of the envelope is one of the driving parameters
 of the explosion, 3) the physics of the core and explosion mechanism
 of all core collapse SNe are not to be fundamentally different, 
regardless of the external appearance of the supernova.

The great observational diversity of CCSNe has not been fully
 understood even if it clearly involves the progenitors masses and
configurations at the time of explosion. Whereas SNII-P are thought
 to originate from isolated massive stars, a generalized scenario
 has been proposed in which common envelope evolution in massive
 binary systems with varying mass ratios and separations of the components
 can lead to various degree of stripping of the envelope
 (Nomoto et al. \cite{nomot1995}).
 According to
 this scenario the sequence of types IIL Ib Ic is ordered according
 to a decreasing mass of the envelope (Turatto et al. \cite{turat2002}).

\section{How to estimate $\Psi(E_{51})$?}

       For this purpose we use the data (Richardson et al. \cite{richa}), where
 a comparative study of the absolute-magnitude distributions of supernovae
 has been done. They used the Asiago Supernova Catalog (ASC), where
 by June, 2001  the number of events had increased to 1910,
 but number of events,
 for which this study is possible  is much smaller. 
 As far as  the SNII types, either the number
 of events for which accurate peak apparent magnitude have been reported
 remains small (SNe Ib, Ic, II-L, IIn) or the intrinsic dispersion in
 the peak absolute magnitude is large (SNII-P). For the absolute-magnitude
 distribution (in blue filter Mb), the authors consider only SNe within 1 Gpc.
 These distributions  for different types of SNe are presented in Fig. 1.

\begin{figure} 
\centering

\includegraphics[width=11cm]{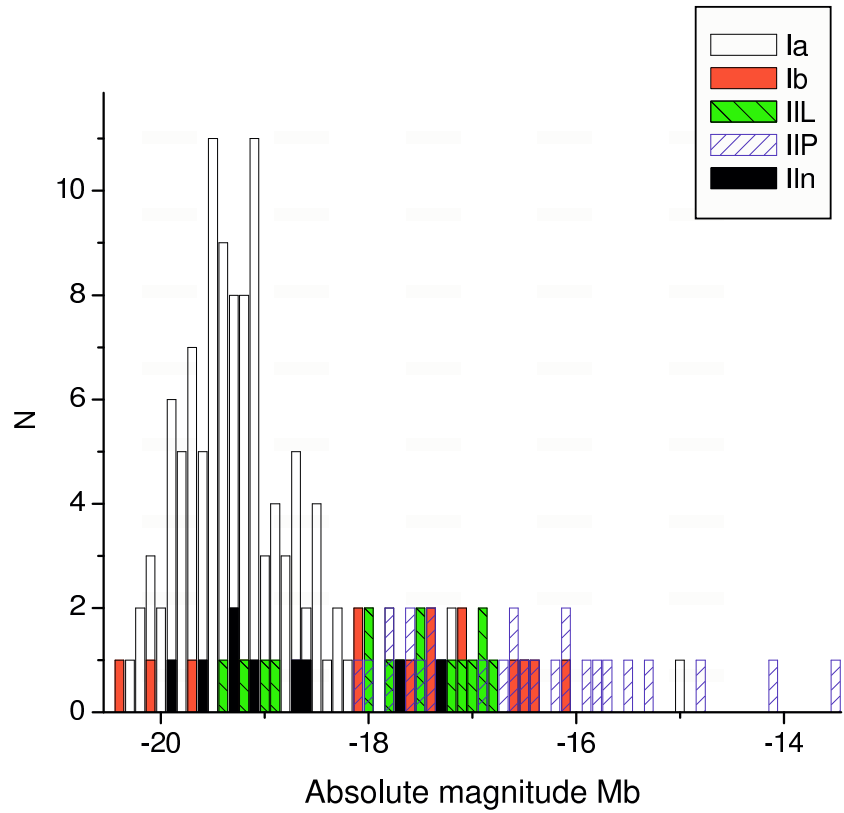}

\caption{ 
Absolute-magnitude distributions of various types of SNe from
 (Richardson et al. \cite{richa}).}
\label{FigVibStab} 

\end{figure}

       The analysis shows that (Richardson et al. \cite{richa}):

 1) At least 7 of 31 SNe in our Galaxy and in galaxies
 within 10 Mpc appear to have been subluminous (Mb$\ge$ -15). Assuming
 that there is observational bias, it appears that more (perhaps much more)
 that 0.2 of all SNe are subluminous, but this fraction remains very uncertain.

 2) Only 20 of 297 extragalactic maximum-light SNe appear to be overluminous
 (Mb$\le$ -20), but it has become clear, that they exist.
 The absolute-magnitude
 dispersion of SNIbc has increased in comparison with the previous works due to
 the discovery of some rather luminous events. The SNe IIn are on average
 the most luminous type of core-collapse SNe.  Considering the strong observational bias in favor of them, 
 it is safe to conclude that the fraction of all SNe that are overluminous
 must be lower than 0.01.

 3) The authors  have approximated
 absolute-magnitude distributions
 for every  types of SNe by Gaussian functions with parameters, listed in
  Table 1.  They  consider  also 'intrinsic' distributions obtained  taking
  into account not only  Galactic extinction, but also  calculating extinction
 distributed for each SN type, averaged over all galaxy inclinations. 
  Moreover,  they divided   Ibc  into two luminosity groups: bright
  and normal. The II-L group was also divided into two groups.

\begin{center}
Table 1: Parameters of Gaussian distributions for 5 (7) main types of SNe from
(Richardson \cite{richa}).
\begin{tabular}{||llll||}
\hline  \hline
SN Type   & $<Mb_{int}> $ & $\sigma _{int}$ & Statistics  \\ \hline  \hline
Normal Ia & -19.46        & 0.56            & 111(30)      \\ \hline
Total  Ibc& -18.04        & 1.39            & 18       \\ \hline
Bright Ibc1& -20.26        & 0.33            & 5         \\ \hline
Normal Ibc2& -17.61        & 0.74            & 13       \\ \hline
Total  IIL& -18.03        & 0.90            & 16      \\ \hline
Bright IIL1& -19.27        & 0.51            & 4         \\ \hline
Normal IIL2& -17.56        & 0.38            & 12       \\ \hline
Total IIP &  -17.00       & 1.12            & 29       \\ \hline
Total IIn &  -19.15       & 0.92            & 9       \\ \hline \hline
\end{tabular}
\end{center}

As was pointed out in (Hamuy \cite{hamuy}),
the physics of the core and explosion mechanism
of all core collapse SNe are not to be fundamentally different, 
so one can expect the presence of correlations
between  average absolute magnitude Mb for given SNe group 
and average energy of explosion for it. At least on the one hand
the tail of most energetic
explosions of Hypernovae is attributed to SNIbc, SNIIn (Hamuy \cite{hamuy}) and
on the other hand they are most luminous (see Table 1). Actually
 thermonuclear explosions of SNIa should be excluded.
So we have made an attemt to estimate the average dependence $E_{51}$(Mb).
  
For this purpose we use calculations of Nadyozhin (Nadyozhin \cite{nadyo}),
where
he makes predictions (on the base of hydrodynamical modelling
of type II plateau supernovae light curves in the frames of the LN85 model
(Litvinova $\&$ Nadyozhin
 \cite{litvi1983}))  for correlations between three observable parameters
(the plateau duration $\Delta t$, the absolute magnitude $M_V$ measured
in V-filter and photospheric
velocity Vph at the middle of the plateau) on the one side and three
physical parameters (the explosion energy $E_{51}$, the mass envelope
expelled $M_{ej}$ and presupernova radius R) on the other side.
According to LN85, the following approximate relations can be used
 to derive $E_{51}$, $M_{ej}$, $R$ from observations.

$$\lg E_{51}=0.135 M_V+2.34 \lg\Delta t+3.13 \lg Vph-4.205$$
$$\lg Mej=0.234 M_V+2.91 \lg\Delta t+1.96 \lg Vph-1.829 \eqno(5)$$
$$\lg R=-0.572 M_V-1.07 \lg\Delta t-2.74 \lg Vph-3.350,$$
where $Mej$ and $R$ are in solar units and $Vph$ in $1000$ km s$^{-1}$.
The analysis of real 14 events SNe II-P shows (Nadyozhin \cite{nadyo}):
for this events the expelled mass, explosion energy and presupernova
radius remain approximately within the limits $Mej \sim 10\div30 M\odot$,
$R\sim 200\div 600 R\odot$, $E_{51} \sim 0.6\div 2.7$.

We rewrite these formulae to exclude the parameter $\Delta t$ and obtain
the following simple relations

$$\lg E_{51}=-0.43 M_V-0.77 \lg R+0.52\lg Mej -5.83 \eqno(6)$$ 
$$\lg Vph=+0.57 \lg E_{51}-0.06 \lg R-0.48\lg Mej+1.32 \eqno(7)$$ 

For appropriate values $Mej$=20, $R$=200, we have 

$$\lg E_{51}=-0.43 M_V-7 \eqno(7)$$ 

And we will use (7) further as a zero approximation. This is a key
dependence for our calculations and the sensitivity of the results
to this dependence  will be discussed in fourth section.

$$\lg Vph=+0.57 \lg E_{51}+0.57 \eqno(8)$$

In formula (1) the maximal energy of accelerated CR depends slowly on
parameters of the explosion $E_{51}$ and $Vph$   as
 $E_{max} \sim 100 (E_{51}\cdot Vph)^{1/3}$ TeV. Using (7), (8), expressions
we obtain  $E_{max}(E_{51})$

$$ \lg E_{max}=0.52\cdot \lg E_{51}-0.02\lg R -0.16\lg Mej +2.43 \eqno(9)$$

For the same appropriate parameters $Mej$=20, $R$=200 we have

$$ \lg E_{max}=0.52*\lg E_{51}+2.1~(TeV) \eqno(9)$$

So we obtain the  dependence

 $$E_{max} = 120 E_{51}^{0.52}~TeV,\eqno(10)$$
 needed for our 
calculation by formula (2). 

For the estimation of  $\Psi(E_{51})$ we use dependence (7) and the sum
of Gaussian distributions in $Mb$ from Table 1 (Richardson al. \cite{richa})
for various
types of SNe with average parameters listed  in Table 1. Since 
$\lg E_{51}$ depends liniary on $Mb$ (7), the distribution in $Mb$ can  
easily be converted to distribution in $dN/d\lg E_{51}$. It will be also 
the sum of Gaussian functions with parameters: 
$<\lg E_{51}>_j=-0.43<Mb>_j-7$,  
$\sigma_j(\lg E_{51})=0.43 \sigma _j(Mb)$, where j means the type of SNe.
Then $dN/d\lg E_{51}$ can be  transformed into $dN/dE_{51}=\Psi(E_{51})$
distribution.
 
     Here it is important to note that for SN Ia the physics of the explosion
is absolutely different. So for SNIa we use
$<E_{51}>\sim 0.8\cdot 10^{51}$ erg  with dispersion of $20\%$.

     The final distribution $\Psi(E_{51})$  for all types of SNe
 is presented in Fig. 2. As can be seen the most of events have
energies $(0.5\div 3\cdot) 10^{51}$ erg. A remarkable peculiarity of the distribution
is a very long tail toward the  high energies up to  $E_{51}$=50. The
reality of this is a key problem of our calculations and will be discussed
in detail in  the last section.

\begin{figure} 
\centering

\includegraphics[width=11cm]{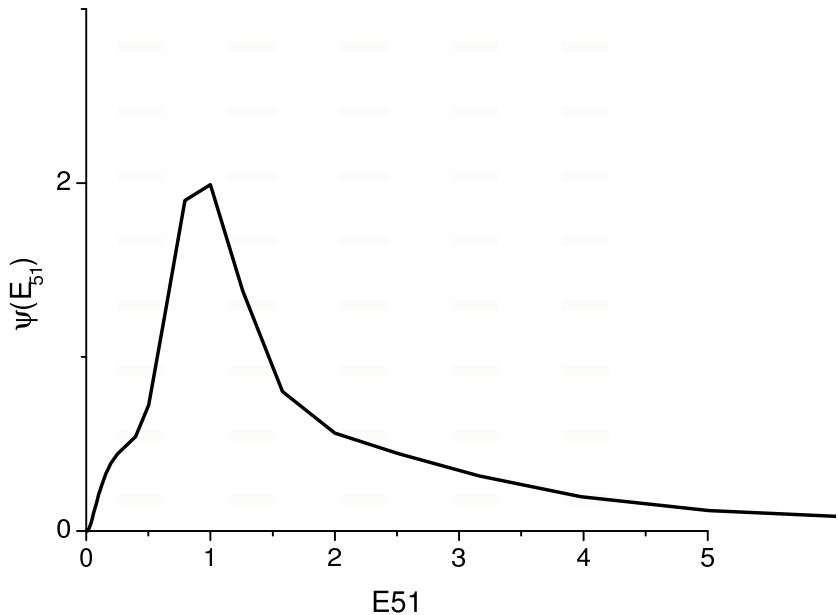}

\caption{ 
 The distribution $\Psi (E_{51})$  for sum of all types of SNe, converted from
  Mb-distribution, presented in Fig.1.}
\label{FigVibStab} 

\end{figure}

      Further we consider the most abundant sites for SNe.
  SN explosions  are not random in the Galaxy, and all of them show
 strong spatial concentration to the centers of galaxies and to the
 arms in spiral galaxies (Bergh \cite{bergh}). The degree of the
 concentration is larger for short-lived massive stars, initiated
 SNII, SNIbc explosions. SNe of type Ia occur in galaxies of all types, even
 far from the place of active star formation zones, due to the long
 evolutionary history in binary systems.   Type Ibc appears only in spiral
 type galaxies and have been associated with a parent population
 of massive stars, perhaps  more massive than SNeII (Turatto \cite{turat2003}),
 and demonstrates even stronger concentration to the centers of galaxies,
 than other types (Bergh \cite{bergh}).

      Numerous regions of very hot and rarefied gas with the
temperature $T=5\cdot 10^6$ K and proton density $n_H=0.003$ cm$^{-3}$ 
occupy $50\%$ of the volume of spiral arms of Galaxy 
(Kononovich $\&$ Moroz \cite{konon}). A source
of heating is thought to be the activity of young stars, first of all
supernovae explosions. So for young massive stars this site of
SNe can be selected as most probable for SNIbc, SNII.

This variant of interstellar medium with parameters
$T=10^6$K, $n_H=0.003$ cm$^{-3}$, magnetic field $B=3\mathrm{\mu G}$ 
was used in (Berezhko \cite{berezh2000b}) being named 'hot phase' 
(Berezinsky et al. \cite{berez}).
 In this variant the maximal predicted energy is $ \sim Z\cdot 300$ TeV.

Numerous regions of neutral gas HI can be divided into
two parts (Kononovich $\&$ Moroz \cite{konon}): the clouds of gas and dust
 with $n_H=10$ cm$^{-3}$
and $T=80$K, that occupy relatively small volume $1\%$, and intercloud
regions, that occupy $50\%$ of the volume of spiral arms with
$n_H=0.1$ cm$^{-3}$ and $T=10^4$ K. These latter  regions were used
in calculations (Berezhko \cite{berezh2000b}) as a most probable site of SNe
with  $T=10^4$K, $n_H=0.3$ cm$^{-3}$,  $B=5\mu$G ('warm phase').
Variant with 'warm phase' gives the maximal energy of acceleration
 about $Z\cdot 100$ TeV. We choose this site as most probable for SNIa.

     Besides that there exists the  temporal correlations resulting from
 the concentration  of the  majority of core collapse SN progenitors
 into OB associations.  The explosion of the first SN among such 
 an association is to be
 followed by several tens of others. This results in the formation of a
 multiple supernova remnants, powered by both the SN explosions and the
 strong winds of Wolf-Rayet stars in the OB associations, which grows as a large
bubble of hot, tenuous plasma known as a superbubble
 SBs(Tomisaka \cite{tomis}; Korpi \cite{korpi}).
                         
 The SB acceleration model has been developed
 by  (Bykov $\&$ Fleishman \cite{bykov1992}), (Bykov $\&$ Uvarov \cite{bykov1999}). 
 They (Bykov $\&$ Toptygin \cite{bykov1997}) estimated the maximal
 energy of accelerated nuclei as  $10^{18}$ eV due to reacceleration effects,
 in the presence of magnetic field
 in the bubble interior of an order of 30 $\mu$G. In this model the spectrum beyond
 the knee is dominated by heavy nuclei.    

        Since SNIIn  explode
in circumstellar medium (in accordance with a definition) 
and can be found in superbubbles, we choose
much stronger  magnetic field $B=30-45\mu$G, that increases the maximal accelerated energy
by a factor of 10-15 for this type of explosions. 

 In the next section we present  numerical results of calculations of all particle 
cosmic ray  spectrum, using all  above  mentioned dependences and parameters
needed for formula (2).

\section{All particles cosmic ray spectrum, numerical results.}

   We divided all nuclei of cosmic rays in 5 rough groups of
 p, He, (C, N, O), (Mg, Si, Ne), Fe with relative intensities  
0.36, 0.25, 0.15, 0.13, 0.15 correspondingly. This chemical composition
takes into account the fact  that heavy components have slightly 
harder spectra beyond 1 TeV, than light ones (Shibata \cite{shiba}),
the contribution of heavy nuclei is increasing
toward higher energies in comparison with
'normal composition', obtained around 1 TeV (Erlykin $\&$ Wolfendale \cite{erlyk2001}).
 The spectrum shapes 
are the same  for all 
components (see formula (3)).

Here it should be noted  that we don't take into account
propagation effects and present source spectra that might be easily
converted to observable spectra in accordance with a standard model 
$\gamma_{sour}=\gamma_{obs}-\Delta \gamma $. 
The value of $\Delta \gamma$  equals to 0.54 (or 0.3 or 0.8)
 depending on propagation model (Jones et al. \cite{jones}).

 Fig. 3  presents the total proton spectrum generated by 7 different 
 types of
 SNe with parameters from Table 1, calculated by formulae (1),(2) with
  $Mb-E_{51}$ conversion function (7), $E_{max}-E_{51}$ dependence (9). 
 Only a portion
 of SNIa  was slightly decreased to $30\%$ (Wheeler \cite{wheel}). 
 The contribution of each SNe type  is presented separately in Fig.~3.

\begin{figure} 
\centering

\includegraphics[width=11cm]{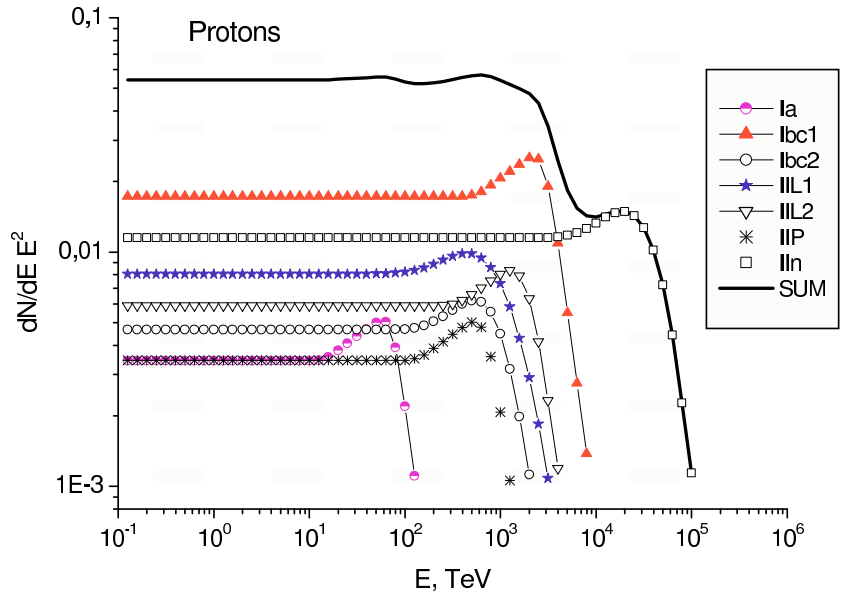}

\caption{ 
 Total proton spectrum (in relative units) generated by 7 different 
 types of SNe with parameters from Table. 1.
 The contribution of each SNe type  is presented separately.
}
\label{FigVibStab} 

\end{figure}

 It can be seen from Fig. 3 that:

 1) The contributions of most energetic explosions are stressed
 significantly due to  expression (4) in our calculations: the total
 number and total energy of accelerated cosmic rays is proportional to the total
 kinetic energy of the explosion. For example, only 5$\%$ of cosmic rays are originated
 by SNIa, while they comprise about  30 $\%$ of all SNe.
 
 2) The  total intensity of CR is practically formed
    by contributions of SNeIbc1 (bright branch, see Table 1) and SNIIn.    

 3) The location of the 'knee' is determined by maximal energy of accelerated
    CRs in the most energetic explosions.

 All obtained regularities can be understood distinctly, if 
 analytical expression for average value of $E_{max}$ is written using
 formula: $E_{max}\sim 300\times E_{51}^{0.5}$ TeV (the value 300 Tev
 corresponds to 'hot phase' of ISM).   The statistical
 weight of $E_{max}$ should be  proportional to the total number
 of accelerated CRs ($\sim E_{51}$ (4).

 $$<E_{max}>=\frac{\int\limits_{E_{51}min}^{E_{51}max} \Psi (E_{51})
 \cdot 300\cdot E_{51}^{0.52} \cdot E_{51}  dE_{51}}
  { \int\limits_{E_{51}min}^{E_{51}max} \Psi (E_{51})
 \cdot  E_{51}  dE_{51}}
=300 \frac{<E_{51}^{1.5}>}{<E_{51}^1>}~ \mathrm{TeV}. \eqno(11)$$

  For a symmetric distribution function $\Psi_j (E_{51})$ with
  a small dispersion,
  as for the case of SNIa, the factor $<E_{51}^{1.5}>~/~<E_{51}^{1}>$
  is close to 1. 
  But for a very asymmetric function, as in Fig. 2, this factor
  can be  
  by many times larger. For example, if we choose power like function
  $\Psi(E_{51})=0.44\cdot E_{51}^{-1.7}$ in interval $E_{51}=1\div 100$ and 
  $\Psi(E_{51})=0.44$ in interval $E_{51}=0.1\div 1$, the value of 
  $<E_{51}^{1.5}>/<E_{51}^{1}>$ is $~5$.
  (This power like shape of $\Psi(E_{51})$ will be discussed in section 5).

 As it can be  seen from Fig. 3, the first 
  'knee' in the proton spectrum is located around 3 PeV, while for the most probable
  energy of explosion $E_{prob}=1$ maximal energy
  $E_{max}\sim 300$ TeV (the variant of
  'hot phase' of interstellar
  medium  in (Berezhko \cite{berezh2000b}).

\begin{figure} 
\centering

\includegraphics[width=11cm]{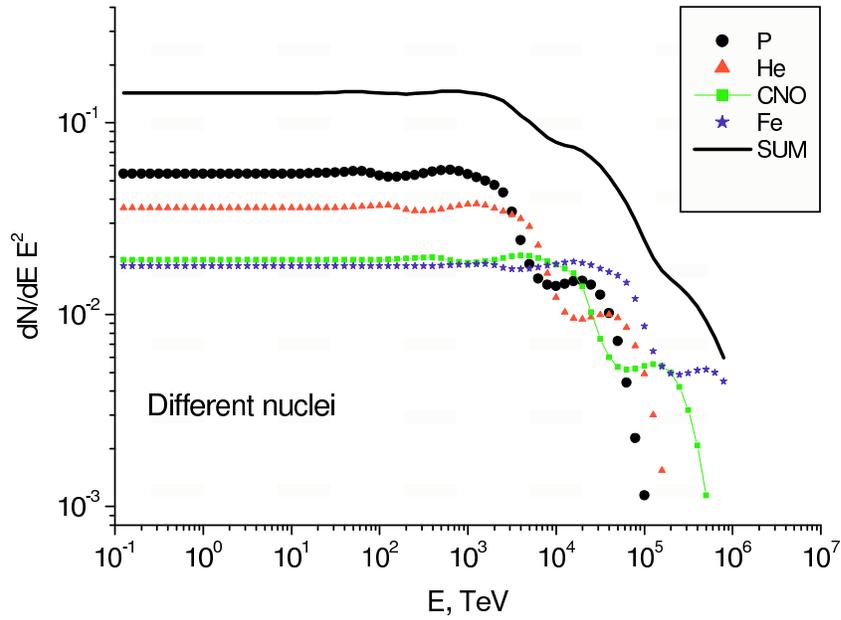}

\caption{ 
  Spectra of  different cosmic ray nuclei.
}
\label{FigVibStab} 

\end{figure}

      The second 'knee' is  formed by the contribution of SNIIn explosions,
  because they are also very energetic (see Table 1) and they explode into
 specific circumstellar medium (in our calculations
 $B=45 \mu G$ for SNIIn).
 
   Fig. 4  presents the spectra of  different cosmic ray nuclei, calculated
  in our model (Mg-Si-Ne group was omitted from the Fig. 4).
  We would like  to remind that $E_{max}(Z)=Z\cdot E_{max}(p)$ (1). 
  Every nuclear component also  has two 'knees'. The maximal energy
  of accelerated CRs is determined by Fe nuclei, originated
  in SNIIn explosions and located around $10^{18}$ eV.

  In Fig. 5 the contributions of various types of SN explosions
  to all particle spectrum are displayed.

\begin{figure} 
\centering

\includegraphics[width=11cm]{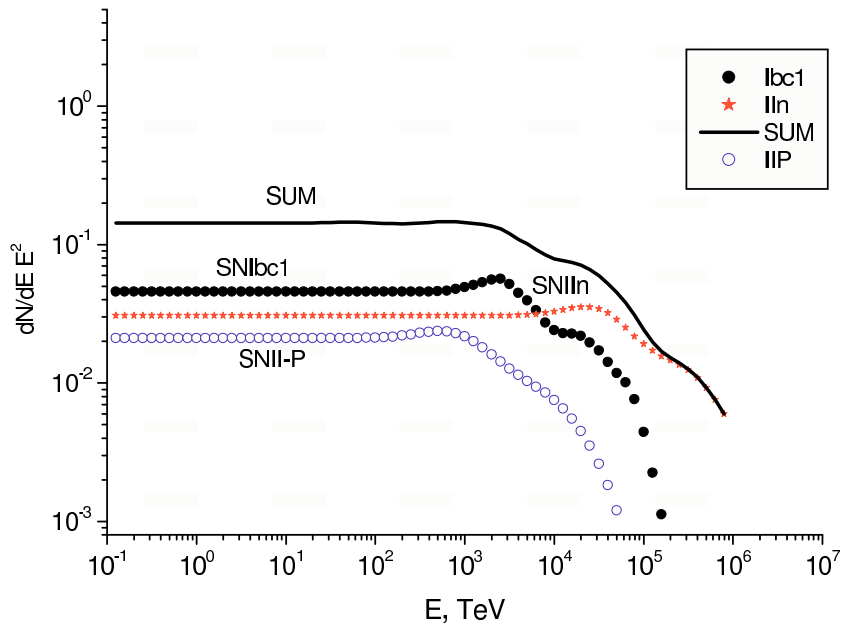}

\caption{ 
 The contributions of various types of SN explosions
  into all particle spectrum.
}
\label{FigVibStab} 

\end{figure}

   The change of a slope of power-like spectrum $\delta \gamma$ beyond
the 'knee' (in the interval 3 PeV to 26$\cdot$ 3 PeV) is determined by a portion of  Fe nuclei $w(Fe)$
in chemical composition of CRs before the 'knee' in the case
when the contribution of SNIIn is  small enough:

$$\delta \gamma= \frac {\lg (1/w(Fe))}{\lg 26}. \eqno(12)$$

For $w(Fe)\sim 0.15-0.20$ the change of the slope is close to $\delta \gamma \sim 0.5$.

If the contribution of SNIIn is noticeable, then  $\delta \gamma$
should be less. Besides, as can be seen from Fig. 5., the more diversity
in explosion energies (as in the case of SNIIP), the more smooth behaviour
of the spectrum beyond the 'knee' is observed. 

The calculations  reproduce on the whole the main features of all particle spectrum,
measured in EAS experiments (Sommers \cite{somme}): the 'knee' around 3~PeV,
the change of slope by $\delta \gamma \sim 0.3-0.5$ beyond the 'knee',
start of dip around $10^{18}$ eV. We do not touch the region beyond
'ankle'.  

 In Fig. 6 the average mass number $<$lnA$>$ in the main variant
of  calculations (when the chemical composition of CRs for SNIIn explosions
is the same as for others) is presented in
comparison with the data obtained in the KASCADE experiment
  (see rev. of (Sommers \cite{somme})). 
The main variant predicts less heavy chemical composition, than in the KASCADE
experiment in the range $10^{16}-10^{17}$ eV.  But in accordance with
(Bykov $\&$ Toptygin \cite{bykov1997}) the supernovae explosions
in superbubbles can generate
the CRs  enriched by heavy nuclei. So a variant when proton and helium
components are absent in CRs generated in superbubbles is also presented
in Fig. 6.
The experimental dependence $<lnA>(E)$ is between the two variants.

\begin{figure} 
\centering

\includegraphics[width=8cm]{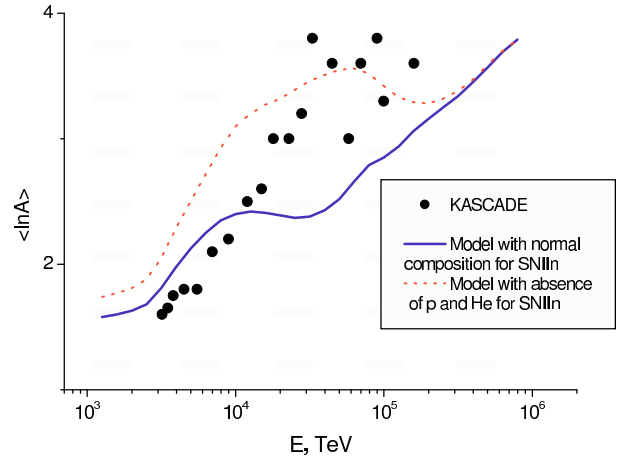}

\caption{ 
 The average mass number $<$lnA$>$ in calculations in
comparison with the data obtained in KASCADE experiment.
}
\label{FigVibStab} 

\end{figure}

\section{The physical interpretation of results}

The key problem in our calculations is a reality of the long tail
in explosion energy distribution $\Psi(E_{51})$ and sensitivity of the 
'knee' location to this tail. To analyze the second item, we show
in Fig. 8 the proton spectrum of CRs with different upper limits
of integrating $E_{51}max$ in formula (2). In the main variant we used
$E_{51}max=80$. Fig. 7 shows that the point of knee location
is determined by the SN explosions with $E_{51}\sim 30\div 60$. 
At $E_{51}max=10$ the 'knee' is around 700 TeV, that is 
higher than 300 TeV due to equation (11): 
$<E_{max}>=300 <E_{51}^{1.5}>,$ but it is not enough to reproduce
experimental data.

\begin{figure} 
\centering

\includegraphics[width=11cm]{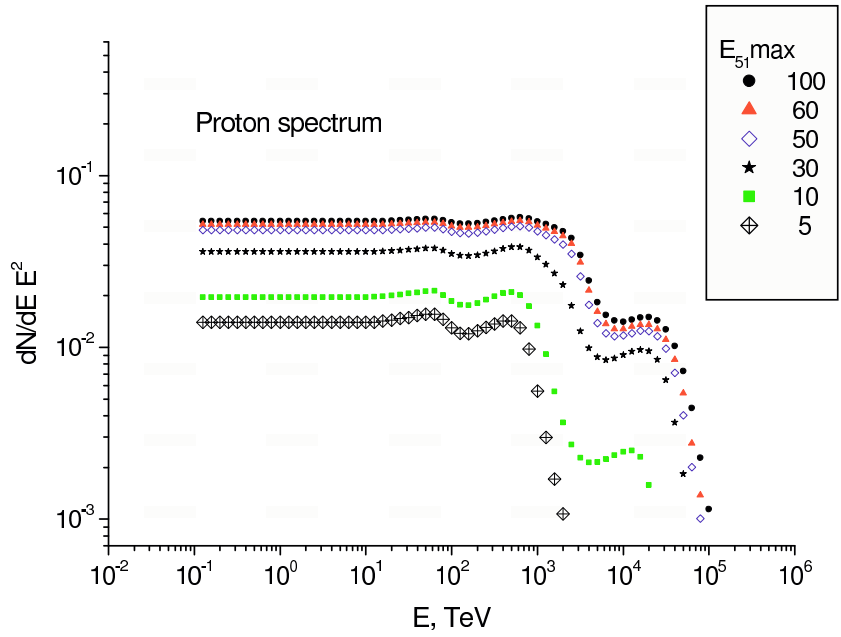}

\caption{ 
 The dependence of 'knee' location on maximal energy of SN explosions.
}
\label{FigVibStab} 

\end{figure}

 One of the most interesting recent developments in the study of SNe
 is the discovery of some very energetic supernovae, whose kinetic energy
 exceeds $10^{52}$ erg - Hypernova (Nomoto et al. \cite{nomot2002}). They can be
 directly identifyed 
 with the explosions determining  cosmic ray 'knee' region not only by energy
 of explosions, but also by the type of core collapse SNe. In our calculations,
 just SNIbc1(bright) and SNIIn give the principal contribution to the formation
 of knee region (see Fig. 3, 5). Among 7 possible Hypernovae 5 have been recognized
 as type Ic (1998bw, 1997ef,1997dq, 1999as, 2002p) and 2 as type IIn
 (1997cy, 1999E) (Nomoto et al. 2002). The Hypernova branch might be
 interpreted as follows. Stars with $M>20-25 M\odot$ form a black hole as
 a compact remnant; whether they become hypernovae or faint SNe may depend
 on the angular momentum in the collapsing core, which, in turn, depends on the
 stellar winds, metallicity, magnetic field and binarity. Hypernovae might
 have rapidly rotating cores owing possibly to the spiraling-in of
 a companion star in a binary system (Nomoto et al. \cite{nomot2002}).

\begin{figure} 
\centering

\includegraphics[width=10cm]{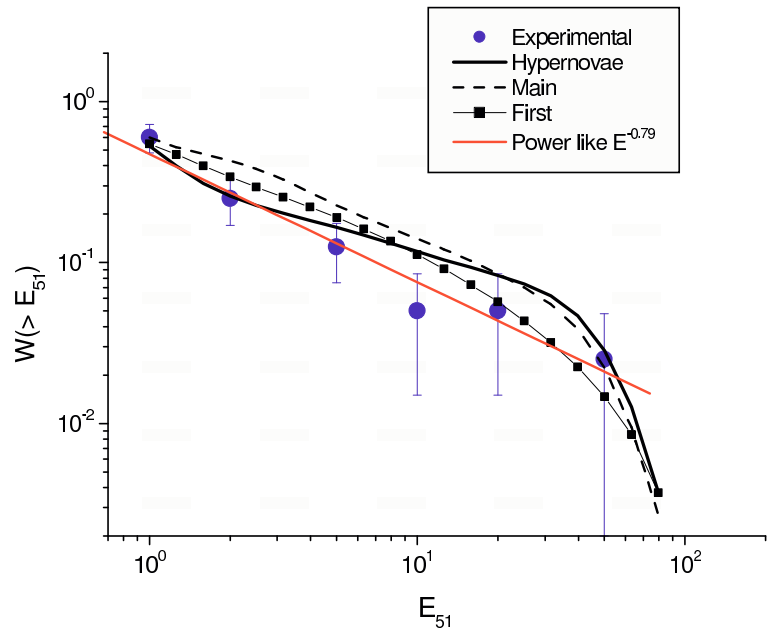}

\caption{ 
 The integral distributions $W(>E_{51})$ for all mentioned in the text
 variants  of calculations.
}
\label{FigVibStab} 

\end{figure}

 Since the physical picture of the Hypernovae explosion   can differ
 from other core collapse SNe, we considered the variant,
 when in key dependence $E_{51}(Mb)$~(6) the parameters of ejected mass and
 radius of progenitor have been chosen different for 
 SNIbc1, SNIIn (see Table 1) and other SNe. In the 'main variant'
 $Mej$=20, $R$=200. The variant with $Mej$=4, $R$=80 for SNIbc1 and SNIIn, but
 $Mej$=10, $R$=600 for other types from Table 1 is called
 'Hypernovae variant' in Fig. 8, Fig. 9.

\begin{figure} 
\centering

\includegraphics[width=11cm]{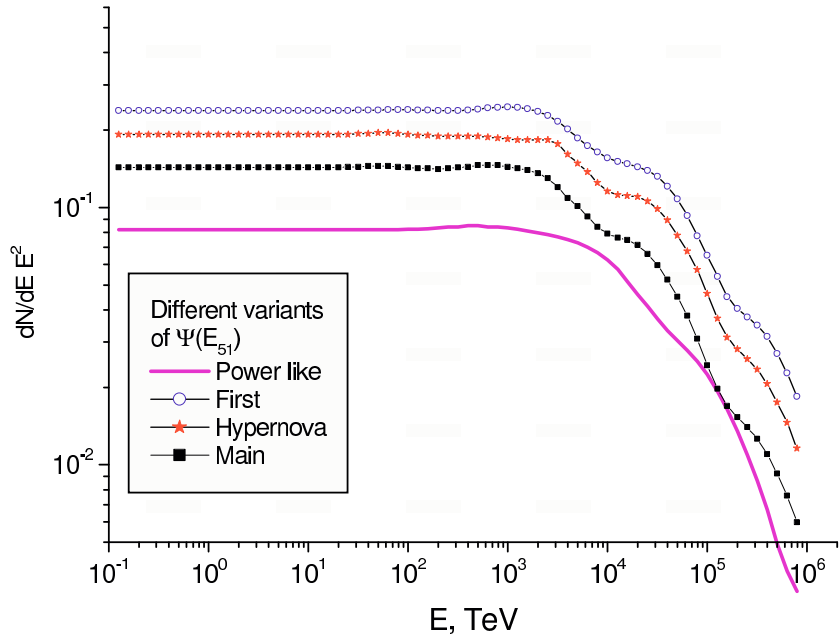}

\caption{ 
 All particle spectra, calculated with different variants
 of  $W(>E_{51})$ presented in Fig.8.
}
\label{FigVibStab} 

\end{figure}

 Besides that, we consider the variant which was historically the first
 to be used in our calculations (the 'first variant'), and  was based on
 a more sophisticated
 estimation of the dependence (6):  $\lg E_{51}=-0.63 Mb-10.6$.
 Here only five groups of SNe from Table 1 without division of SNIbc
 and SNIIL into bright and normal classes were considered.

 In ideal case  the experimental distribution
  $\Psi(E_{51})$ should be used.  As a zero approximation to this case
 we consider of the 
 tables from (Hamuy, \cite{hamuy}) (where    the physical
 parameters for  types II, Ibc, IIdw  real supernovae are presented) and
 constructed the integral distribution $W(>E_{51})$, since our calculation
 is sensitive mainly to the long tail in energies. This distribution
 (see Fig. 8) is
 called 'experimental variant', while in reality it depends on the
  basic theoretical premises and moreover it can be slightly
 distorted by the selection bias.

As it can be seen from Fig. 8, the distribution
 has a power-like shape with the slope of -0.79. It means that differential
 distribution is $\sim E^{-1.79}$. The portion of events with
 $E_{51}>1$ was normalized to 0.6. In the region of $E_{51}=0.1-1$ we choose
 $N(E_{51})=const$.

 The integral distributions $W(>E_{51})$ for the above mentioned variants
 of calculations are presented in Fig. 8.  In Fig. 9 we show
 all particle CR spectrum calculated for every variant of $W(>E_{51})$.

   The contribution of events with  energy  $E_{51}\sim 50$
   comes to $\sim 2\%$ in all cases, but the contribution
   of events around $E_{51}\sim 10$  is  by a factor of 2-3 larger than in
   'experimental'  distribution. But this difference occured to be not
   crucial for all particle CR spectra presented in Fig. 9, because
   the knee is mostly sensitive just to the energy of explosion
   $E_{51}\sim 30-50$, as  shown in Fig. 7.

       So one can conclude, that the portion of events responsible
for the formation of the knee in the CR spectrum  comes to $\sim 2\pm 1\% $ meaning
that  with a usual SN-rate 
of about 0.01 year$^{-1}$ the Hypernova rate is about 0.0002 year$^{-1}$.
If the total power of 100 SNe with average energy $E_{51}=1$ in our Galaxy is enough
to provide  the total energy of Galactic cosmic rays 
 $\sim 10^{-12}$ erg/cm$^3$ (Berezinsky et al \cite{berez}), then
2 SNe with $E_{51}=50$ or 3 SNe  with $E_{51}=30$ can also provide 
the total energy of Galactic cosmic rays. 

If the time of CR life  in our Galaxy is about $3\cdot 10^7$ year 
and  Hypervae rate is 0.0002 year$^{-1}$, it means that about 
 $6\cdot 10^3$ explosions
provide the CR intensity in our Galaxy. But for high energy CRs (around
1 PeV) the time of CR life $T_{esc}$  is much smaller due to  decreasing of
escape lenght  $\lambda_{esc}$   as $E^{-\alpha}$
 ($\alpha=0.54$)  beyond the energy $5$ GeV (Jones et al. \cite{jones}).
The number of  explosions giving the dominating input to CRs around
the $E \sim 1$ PeV  might be rather small ($\sim 10-15$) in hole Galaxy
and only several explosions in the nearby region.

 During the last millennium 5 SNe are known
to have exploded (1006, 1054, 1572, 1054, 1679(Cas A)),
 see review of (Weeler \cite{wheel}).
 The supernova of
 1006 was the brightest ever recorded, brighter than Venus and perhaps as
 bright as a quarter Moon. Of these events, only
 SN 1054 produced an observable compact remnant, the Crab pulsar. Although
 there is still some controversy, the events of 1006, 1572, 1604 are 
 generally thought  not to have left any compact remnant, but to have resulted in
 complete disruption of the star (Weeler \cite{wheel}).

\section{Conclusions}

\begin{enumerate}

  \item The main idea of this work  to take into account the
 distribution of
Supernovae in types and explosion energies results in a very promising
conclusion: the maximal energy of accelerated CRs moves to higher energies
by a factor of 5-10 in comparison with the value 100-300 TeV, predicted by standard
model of CR acceleration, due to diversity
in SN explosion energies. 
This conclusion is based essentially on the premise that the energy of
 accelerated CRs
is proportional to the energy of an explosion, that  stresses
the input of high energy explosions and suppresses the contribution
of low energy explosions to the total CR flux. It is worth to note here  that
rather  significant diversity in energies (around one order) exists even inside the 
SNIIP group   (Nadyozhin \cite{nadyo}; Hamuy \cite{hamuy}).

\item The knee in CR spectrum can  quantitatively be explained by the dominating
 contribution of SNIbc1, SNIIn  explosions with energy
 of ($\sim 30-50)\cdot 10^{51}$ erg, that might be identified with Hypernovae.

\item The estimated rate of these energetic explosions is  about 1-3 per
$10^4$ years being  enough to provide the total power of Galactic cosmic rays.
 As a whole about 6000 stars provide the CR flux
 at lower energies,
but may be only several explosions give the dominating contribution to the CR
 flux around the knee region (nearby of local Solar system).

\end{enumerate}

The latter conclusion may intersect with the idea, proposed by
 Erlykin and Wol\-fen\-dale (Erlykin $\&$ Wolfendale \cite{erlyk1997}),
 that a single  nearby local SNR  accelerates the particles
 and gives the  dominating input (mainly by O and Fe nuclei)
 to the knee region. But the CR propagation in the Galaxy
  is absent for this SNR. In our model the most energetic
 explosions give the dominating contribution to the whole
 energy range of CRs and their  propagation in Galaxy
  should be taken into account,   while we  do not consider here
these effects. Fractal diffusion seems likely to be very usefull
to help  CRs to reach the Solar system (Lagutin et al. \cite{lagut}).

\vskip 3mm

{\bf Acknowledgements.}
I acknowledge numerous discussions on this problem in private conversations
and at seminars with many physicists: prof. K.A.Postnov, prof. D.Yu.Tsvetkov,
prof. V.S.Ptuskin, prof. A.D.Erlykin, prof. G.T.Zatsepin, prof. O.G.Ryazhkaya,
prof. M.I.Panasyuk,
prof. N.N.Kalmykov, prof. T.M. Roganova, L.A.Kuzmichev, K.A.Managadze,
N.V.Sokolskaya.
I am very thankful to all
members of the RUNJOB and NUCLEON collaborations, where I am engaged,
for discussions and
support.

\end{document}